# Charmonium at CLEO-c

**David H Miller**[*†]
*Purdue University,West Lafayette, USA*
*E-mail:* miller@physics.purdue.edu

The charmonium results presented in this paper are part of a continuing program using the CLEO-c detector to produce high precision results on both open charm decays and charmonium systems [1]. The results include:
Observation of the $h_c(^1P_1)$
Branching fractions for $\mathcal{B}(J/\psi \to e^+e^-)$ and $\mathcal{B}(J/\psi \to \mu^+\mu^-)$
Observation of $\psi(3770) \to \pi\pi J/\psi$ and Measurement of $\Gamma_{ee}[\psi(2S)]$
Branching Fractions for $\psi(2S)$-to-$J/\psi$ Transitions
First Observation of $\psi(3770) \to \gamma\chi_{c1} \to \gamma\gamma J/\psi$
Two Photon Width of $\chi_{c2}$
Hadronic decays of the $\psi(2S)$

[*] Speaker.
[†] I would like to thank my CLEO colleagues and the CESR staff

## 1. Charmonium results from the $\psi(2S)$ and the $\psi(3770)$ [2]

The focus of the current CLEO-c program is on high precision measurements of charm physics, both open charm and charmonium bound states, from data taken at the $\psi(3770)$, $\psi(2S)$ and above $D_s\bar{D}_s$ threshold [1]. In addition the previous detector,CLEO III, accumulated data at the $\psi(2S)$. The results presented in this paper come from 5.85 $pb^{-1}$ taken at the $\psi(2S)$, 20.46 $pb^{-1}$ of continuum taken 50 MeV below the $\psi(2S)$ and 281$pb^{-1}$ at the $\psi(3770)$.This paper summarizes the results and detailed description of each analysis can be found in the references.

### 1.1 Observation of the $h_c(^1P_1)$ [3]

The $h_c(^1P_1)$ state of charmonium has been observed in the isospin-violating reaction

$$e^+e^- \to \psi(2S) \to \pi^0 h_c\ ,\ h_c \to \gamma\eta_c\ ,\ \pi^0 \to \gamma\gamma. \tag{1.1}$$

in which the $\eta_c$ decays are measured exclusively or inclusively. In the exclusive analysis, $\eta_c$ are reconstructed in seven channels: $K^0_S K^\pm\pi^\mp$, $K^0_L K^\pm\pi^\mp$, $K^+K^-\pi^+\pi^-$, $\pi^+\pi^-\pi^+\pi^-$, $K^+K^-\pi^0$, $\pi^+\pi^-\eta(\to\gamma\gamma)$, and $\pi^+\pi^-\eta(\to\pi^+\pi^-\pi^0)$. The sum of the branching fractions is $(9.7\pm2.7)\%$ [4]. These measurements allow a precise determination of the mass of $h_c$ and the branching fraction product $\mathcal{B}_\psi\mathcal{B}_h$, where $\mathcal{B}_\psi \equiv \mathcal{B}(\psi(2S)\to\pi^0 h_c)$ and $B_h \equiv \mathcal{B}(h_c\to\gamma\eta_c)$. The results are combined to obtain $M(h_c) = 3524.4\pm0.6\pm0.4$ MeV and $\mathcal{B}(\psi(2S)\to\pi^0 h_c)\times\mathcal{B}(h_c\to\gamma\eta_c) = (4.0\pm0.8\pm 0.7)\times10^{-4}$. and the hyperfine splitting is:

$$\Delta M_{hf}(\langle M(^3P_J)\rangle - M(^1P_1)) = +1.0\pm0.6\pm0.4\,\mathrm{MeV}.$$

The combined result for $M(h_c)$ is consistent with the spin-weighted average of the $\chi_{cJ}$ states.

### 1.2 Branching fractions for $\mathcal{B}(J/\psi\to e^+e^-)$ and $\mathcal{B}(J/\psi\to\mu^+\mu^-)$ [5]

The measurements of $\mathcal{B}(J/\psi\to e^+e^-)$ and $\mathcal{B}(J/\psi\to\mu^+\mu^-)$ are performed using the decay $\psi(2S)\to\pi^+\pi^- J/\psi$. The experimental procedure is straightforward and consists of determining the ratios of the numbers of exclusive $J/\psi\to\ell^+\ell^-$ decays for $\ell = e$ and $\mu$, $N_{e^+e^-}$ and $N_{\mu^+\mu^-}$, to the number of inclusive $J/\psi\to X$ decays, $N_X$, where $X$ means all final states. We obtain $\mathcal{B}(J/\psi\to e^+e^-) = (5.945\pm0.067\pm0.042)\%$ and $\mathcal{B}(J/\psi\to\mu^+\mu^-) = (5.960\pm0.065\pm0.050)\%$, leading to an average of $\mathcal{B}(J/\psi\to\ell^+\ell^-) = (5.953\pm0.056\pm0.042)\%$ and a ratio of $\mathcal{B}(J/\psi\to e^+e^-)/\mathcal{B}(J/\psi\to\mu^+\mu^-) = (99.7\pm1.2\pm0.6)\%$, all consistent with, but more precise than, previous measurements.

### 1.3 Observation of $\psi(3770)\to\pi\pi J/\psi$ [6]

Using the decays $\psi(3770)\to XJ/\psi$, $X = \pi^+\pi^-$ (13$\sigma$ significance) and $\pi^0\pi^0$ (3.8$\sigma$) the following branching fractions are obtained: $\mathcal{B}(\psi(3770)\to\pi^+\pi^- J/\psi) = (214\pm25\pm22)\times10^{-5}$ and $\mathcal{B}(\psi(3770)\to\pi^0\pi^0 J/\psi) = (97\pm35\pm20)\times10^{-5}$. The radiative return process $e^+e^-\to\gamma\psi(2S)$ populates the same event sample and is used to measure $\Gamma_{ee}[\psi(2S)] = (2125\pm26\pm82)$ eV.

### 1.4 Branching Fractions for $\psi(2S)$-to-$J/\psi$ Transitions [9]

New measurements have been made of the inclusive and exclusive branching fractions for $\psi(2S)$. which are either the most precise measurements to date or the first direct measurements. These results are shown in Table 1.

**Table 1:** For each mode: The detection efficiency, $\epsilon$, in percent; the numbers of events found in the $\psi(2S)$ and continuum samples, $N_{\psi(2S)}$ and $N_{\rm cont}$; the number of $\psi(2S)$ related background events, $N_{\rm bgd}$; the branching fraction in percent and its ratio to $\mathcal{B}_{X\,J/\psi}$ and $\mathcal{B}_{\pi^+\pi^-\,J/\psi}$, also in percent.

| Channel | $\epsilon$ | $N_{\psi(2S)}$ | $N_{\rm cont}$ | $N_{\rm bgd}$ | $\mathcal{B}$ | $\mathcal{B}/\mathcal{B}_{X\,J/\psi}$ | $\mathcal{B}/\mathcal{B}_{\pi^+\pi^-\,J/\psi}$ |
|---|---|---|---|---|---|---|---|
| $\pi^+\pi^- J/\psi$ | 49.3 | 60344 | 221 | 113 | $33.54 \pm 0.14 \pm 1.10$ | $56.37 \pm 0.27 \pm 0.46$ | |
| $\pi^0\pi^0 J/\psi$ | 22.2 | 13399 | 67 | 115 | $16.52 \pm 0.14 \pm 0.58$ | $27.76 \pm 0.25 \pm 0.43$ | $49.24 \pm 0.47 \pm 0.86$ |
| $\eta J/\psi$ | 22.6 | 2793 | 17 | 116 | $3.25 \pm 0.06 \pm 0.11$ | $5.46 \pm 0.10 \pm 0.07$ | $9.68 \pm 0.19 \pm 0.13$ |
| $\eta(\to\gamma\gamma)J/\psi$ | 16.9 | 2065 | 14 | 103 | $3.21 \pm 0.07 \pm 0.11$ | $5.39 \pm 0.12 \pm 0.06$ | $9.56 \pm 0.21 \pm 0.14$ |
| $\eta(\to\pi^+\pi^-\pi^0)J/\psi$ | 5.8 | 728 | 3 | 13 | $3.39 \pm 0.13 \pm 0.13$ | $5.70 \pm 0.21 \pm 0.13$ | $10.10 \pm 0.38 \pm 0.22$ |
| $\pi^0 J/\psi$ | 13.9 | 88 | 3 | 20 | $0.13 \pm 0.01 \pm 0.01$ | $0.22 \pm 0.02 \pm 0.01$ | $0.39 \pm 0.04 \pm 0.01$ |
| $\gamma\chi_{c0}\to\gamma\gamma J/\psi$ | 23.4 | 172 | 20 | 17 | $0.18 \pm 0.01 \pm 0.02$ | $0.31 \pm 0.02 \pm 0.03$ | $0.55 \pm 0.04 \pm 0.06$ |
| $\gamma\chi_{c1}\to\gamma\gamma J/\psi$ | 30.6 | 3688 | 46 | 21 | $3.44 \pm 0.06 \pm 0.13$ | $5.77 \pm 0.10 \pm 0.12$ | $10.24 \pm 0.17 \pm 0.23$ |
| $\gamma\chi_{c2}\to\gamma\gamma J/\psi$ | 28.6 | 1915 | 56 | 62 | $1.85 \pm 0.04 \pm 0.07$ | $3.11 \pm 0.07 \pm 0.07$ | $5.52 \pm 0.13 \pm 0.13$ |
| $X J/\psi$ | 65.3 | 151138 | 37916 | 123 | $59.50 \pm 0.15 \pm 1.90$ | | |

### 1.5 First Observation of $\psi(3770)\to\gamma\chi_{c1}\to\gamma\gamma J/\psi$ [10]

The non-$D\bar{D}$ decay $\psi(3770)\to\gamma\chi_{c1}$ is observed. The two-photon cascades to $J/\psi$ and $J/\psi\to\ell^+\ell^-$ are analyzed and the results are: $\sigma(e^+e^-\to\psi(3770))\times\mathcal{B}(\psi(3770)\to\gamma\chi_{c1}) = (20.4\pm3.7\pm2.4)$ pb and branching fraction $\mathcal{B}(\psi(3770)\to\gamma\chi_{c1}) = (3.2\pm0.6\pm0.4)\times10^{-3}$. The 90% C.L. upper limits for the transition to $\chi_{c2}$ ($\chi_{c0}$) : $\sigma\times\mathcal{B} < 10.8$ pb ($<295$ pb) and $\mathcal{B} < 1.7\times10^{-3}$ ($<46\times10^{-3}$).

### 1.6 Two Photon Width of $\chi_{c2}$ [11]

A new measurement has been made of the two-photon width of $\chi_{c2}$ using reaction

$$e^+e^-\to e^+e^-(\gamma\gamma)\,,\ \gamma\gamma\to\chi_{c2}\to\gamma J/\psi\to\gamma l^+l^-. \tag{1.2}$$

The results are $\Gamma_{\gamma\gamma}(\chi_{c2})\mathcal{B}(\chi_{c2}\to\gamma J/\psi)\mathcal{B}(J/\psi\to e^+e^-+\mu^+\mu^-) = 13.2\pm1.4({\rm stat})\pm1.1({\rm syst})$ eV, and $\Gamma_{\gamma\gamma}(\chi_{c2}) = 559\pm57({\rm stat})\pm48({\rm syst})\pm36({\rm br})$ eV. This result is in excellent agreement with the result of two-photon fusion measurement by Belle [7] and also the $\bar{p}p\to\chi_{c2}\to\gamma\gamma$ measurement [8], when they are both reevaluated using the recent CLEO result for the radiative decay $\chi_{c2}\to\gamma J/\psi$.

### 1.7 Hadronic decays of the $\psi(2S)$ [12]

The states $J/\psi$ and $\psi(2S)$ are non-relativistic bound states of a charm and an anti-charm quark. In perturbative QCD the decays of these states are expected to be dominated by the annihilation of the constituent $c\bar{c}$ into three gluons or a virtual photon. The partial width for the decays into an exclusive hadronic state $h$ is expected to be proportional to the square of the $c\bar{c}$ wave function overlap at zero quark separation, which is well determined from the leptonic width [4]. Since the strong coupling constant, $\alpha_s$, is not very different at the $J/\psi$ and $\psi(2S)$ masses, it is expected that for any state $h$ the $J/\psi$ and $\psi(2S)$ branching ratios are related by:

$$Q_h = \frac{\mathcal{B}(\psi(2S)\to h)}{\mathcal{B}(J/\psi\to h)} \approx \frac{\mathcal{B}(\psi(2S)\to\ell^+\ell^-)}{\mathcal{B}(J/\psi\to\ell^+\ell^-)} = (12.7\pm0.5)\%, \tag{1.3}$$

where $\mathcal{B}$ denotes a branching fraction, and the leptonic branching fractions are taken from the Particle Data Group (PDG) [4]. This relation is sometimes called "the 12% rule". The results for a wide variety of mesonic and baronic decays with and without strange particles are shown in tables 2 and 3.

**Table 2:** For each final state $h$ the following quantities are given: the decay mode, the number of events attributable to $\psi(2S)$ decay, $N_S$, the average efficiency, $\epsilon$; the absolute branching fraction with statistical (68% C.L.) and systematic errors; previous branching fraction measurements from the PDG [4], and the $Q_h$ value. For $\eta 3\pi$, the two decays modes $\eta 3\pi(\eta \to \gamma\gamma)$ and $\eta 3\pi(\eta \to 3\pi)$ are combined on line $\eta 3\pi$.

| mode $h$ | $N_S$ | $\varepsilon$ | $\mathcal{B}(\psi(2S) \to h)$ (units of $10^{-4}$) | $\mathcal{B}$ (PDG) (units of $10^{-4}$) | $Q_h$ (%) |
|---|---|---|---|---|---|
| $2(\pi^+\pi^-)$ | 308.0 | 0.4507 | 2.2±0.2±0.2 | 4.50±1.00 | 5.55±1.53 |
| $\rho\pi^+\pi^-$ | 285.5 | 0.4679 | 2.0±0.2±0.4 | 4.20±1.50 | - |
| $2(\pi^+\pi^-)\pi^0$ | 1702.6 | 0.2115 | 26.1±0.7±3.0 | 30.00±8.00 | 7.76±1.10 |
| $\eta\pi^+\pi^-$ | 7.2 | 0.0416 | $< 1.6$ | - | - |
| $\omega\pi^+\pi^-$ | 391.0 | 0.1553 | 8.2±0.5±0.7 | 4.80±0.90 | 11.35±1.94 |
| $\eta 3\pi(\eta \to \gamma\gamma)$ | 201.7 | 0.0639 | 10.3±0.8±1.4 | - | - |
| $\eta 3\pi(\eta \to 3\pi)$ | 50.0 | 0.0199 | 8.1±1.4±1.6 | - | - |
| $\eta 3\pi$ | | | 9.5±0.7±1.5 | - | - |
| $\eta' 3\pi$ | 12.8 | 0.0092 | 4.5±1.6±1.3 | - | - |
| $K^+K^-\pi^+\pi^-$ | 817.2 | 0.3742 | 7.1±0.3±0.4 | 16.00±4.00 | 9.85±3.23 |
| $\rho K^+K^-$ | 223.8 | 0.3361 | 2.2±0.2±0.4 | - | - |
| $\phi\pi^+\pi^-$ | 47.6 | 0.1744 | 0.9±0.2±0.1 | 1.50±0.28 | 11.07±3.30 |
| $K^+K^-\pi^+\pi^-\pi^0$ | 711.6 | 0.1818 | 12.7±0.5±1.0 | - | 10.59±2.81 |
| $\eta K^+K^-$ | 4.3 | 0.0354 | $< 1.3$ | - | - |
| $\omega K^+K^-$ | 76.8 | 0.1288 | 1.9±0.3±0.3 | 1.50±0.40 | 10.19±2.96 |
| $2(K^+K^-)$ | 59.2 | 0.3118 | 0.6±0.1±0.1 | - | 6.71±2.74 |
| $\phi K^+K^-$ | 36.8 | 0.1511 | 0.8±0.2±0.1 | 0.60±0.22 | 5.14±1.53 |
| $2(K^+K^-)\pi^0$ | 44.7 | 0.1339 | 1.1±0.2±0.2 | - | - |
| $p\bar{p}\pi^+\pi^-$ | 904.5 | 0.4943 | 5.9±0.2±0.4 | 8.00±2.00 | 9.90±1.16 |
| $\rho p\bar{p}$ | 61.1 | 0.4119 | 0.5±0.1±0.2 | - | - |
| $p\bar{p}\pi^+\pi^-\pi^0$ | 434.9 | 0.1921 | 7.3±0.4±0.6 | - | 18.70±5.80 |
| $\eta p\bar{p}$ | 9.8 | 0.0399 | 0.8±0.3±0.3 | - | 3.80±2.09 |
| $\omega p\bar{p}$ | 21.2 | 0.1129 | 0.6±0.2±0.2 | 0.80±0.32 | 4.69±2.22 |
| $p\bar{p}K^+K^-$ | 30.1 | 0.3671 | 0.3±0.1±0.0 | - | - |
| $\phi p\bar{p}$ | 4.3 | 0.1732 | $< 0.24$ | $< 0.26$ | - |
| $\Lambda\bar{\Lambda}\pi^+\pi^-$ | 73.4 | 0.0844 | 2.8±0.4±0.5 | - | - |
| $\Lambda\bar{p}K^+$ | 74.0 | 0.2472 | 1.0±0.1±0.1 | - | 10.92±2.93 |
| $\Lambda\bar{p}K^+\pi^+\pi^-$ | 45.8 | 0.0847 | 1.8±0.3±0.3 | - | - |

**Table 3:** Branching ratios of $\psi(2S)$ decaying to baryon-antibaryon pairs. The last column shows the background subtracted continuum cross-section..

| Modes<br>$\epsilon$ | $S_{\psi(2S)}$<br>$\mathcal{B}(10^{-4})$ | $B_{\psi(2S)}$<br>Q(%) | $f_S \cdot B_c$<br>$\sigma_{cont}(pb)$ | $B_{\mathrm{x}f}$ |
|---|---|---|---|---|
| $p\overline{p}$ | 66.6% | 2.87±0.12±0.15 | 13.6±1.1 | 1.5±0.37±0.13 |
| $\Lambda\overline{\Lambda}$ | 20.1% | 3.28±0.23±0.25 | 25.2±3.5 | <2.0 @90 CL |
| $\Sigma^+\overline{\Sigma^+}$ | 4.1% | 2.57±0.44±0.68 | - | - |
| $\Sigma^0\overline{\Sigma^0}$ | 7.2% | 2.63±0.35±0.21 | 20.7±4.2 | - |
| $\Xi^-\overline{\Xi^-}$ | 8.6% | 2.38±0.30±0.21 | 13.2±2.2 | <3.5 @90 CL |
| $\Xi^0\overline{\Xi^0}$ | 2.4% | 2.75±0.64±0.61 | | <14 @90 CL |
| $\Xi^{*0}\overline{\Xi^{*0}}$ | 0.6% | $0.72^{+1.48}_{-0.62} \pm 0.10$<br>(<3.2 @90 CL) | - | - |
| $\Omega^-\overline{\Omega^-}$ | 1.9% | $0.70^{+0.55}_{-0.33} \pm 0.10$<br>(<1.6 @90 CL) | - | - |

## References

[1] R. A. Briere *et al.*,Cornell Report CLNS 01/1742 (2001)

[2] These results were presented at the International Europhysics Conference on High Energy Physics, July 21st - 27th, 2005, Lisboa, Portugal.

[3] J.L.Rosner *et al.*, Phys. Rev. Lett.**95**:102003,2005 [arXiv: hep-ex/0505073 (2005)]

[4] Particle Data Group S. Eidelman *et al.*, Phys. Lett. B **592**, 1 (2004)

[5] Z. Li *et al.*, Phys. Rev. **D71**:111103,2005[arXiv: hep-ex/0503027 (2005)]

[6] N. E. Adam *et al.*,(submitted to Phys. Rev. Lett.)(arXiv: hep-ex/0508023 (2005)]

[7] K. Abe *et al.*, Phys. Lett. **B 540**, 33 (2002)

[8] M. Ambrogiani *et al.*, Phys. Rev. **D 62**, 052002 (2000)

[9] N. E. Adam *et al.*, Phys. Rev. Lett. **94**:232002,2005 '[arXiv: hep-ex/0503028 (2005)]

[10] T. E. Coan *et al.*,(submitted to Phys. Rev. Lett.), (arXiv: hep-ex/0509030 (2005))

[11] S. Dobbs *et al.*,(submitted to Phys. Rev. Lett.), (arXiv: hep-ex/0510033 (2005))

[12] R. A. Briere *et al.*, Phys. Rev. Lett.**95**:062001,2005[arXiv: hep-ex/050101 (2005)], T. K. Pedlar *et al.*,(submitted to Phys. Rev. Lett.), (arXiv: hep-ex/0505057 (2005))